\documentclass[lettersize,journal]{IEEEtran}
\usepackage{amsmath,amsfonts}
\usepackage{algorithmic}
\usepackage{algorithm}
\usepackage{array}
\usepackage[caption=false,font=normalsize,labelfont=sf,textfont=sf]{subfig}
\usepackage{textcomp}
\usepackage{stfloats}
\usepackage{url}
\usepackage{verbatim}
\usepackage{graphicx}
\usepackage{cite}
\usepackage{booktabs} 
\usepackage{multirow}
\usepackage{xcolor}
\usepackage{hyperref}
\usepackage{booktabs}

\begin{document}

\title{Ev-Layout: A Large-scale Event-based Multi-modal Dataset for Indoor Layout Estimation and Tracking}

\author{
    Xucheng Guo, Yiran Shen\IEEEauthorrefmark{1},~\IEEEmembership{Senior Member, IEEE}, Xiaofang Xiao, \\
    Yuanfeng Zhou, and Lin Wang\IEEEauthorrefmark{1}%
    \thanks{\IEEEauthorrefmark{1}Corresponding authors: Yiran Shen and Lin Wang.}%
    \IEEEcompsocitemizethanks{
        \IEEEcompsocthanksitem Xucheng Guo, Yiran Shen, Xiaofang Xiao, and Yuanfeng Zhou are with the School of Software, Shandong University, China.\protect\\
        E-mail: xucheng.guo@mail.sdu.edu.cn, \{yiran.shen; yfzhou\}@sdu.edu.cn
        \IEEEcompsocthanksitem Lin Wang is with the School of Electrical and Electronic Engineering, Nanyang Technological University, Singapore.\protect\\
        E-mail: linwang@ntu.edu.sg
    }
}

\markboth{Journal of \LaTeX\ Class Files,~Vol.~14, No.~8, August~2021}%
{Shell \MakeLowercase{\textit{et al.}}: A Sample Article Using IEEEtran.cls for IEEE Journals}


\maketitle

\begin{abstract}

This paper presents \textbf{Ev-Layout}, a novel large-scale event-based multi-modal dataset designed for indoor layout estimation and tracking. 
Ev-Layout brings pivotal contributions to the community by \textbf{\textit{1)}} utilizing a hybrid data collection platform (with head-mounted display and VR interface) -- integrating both RGB and bio-inspired event cameras -- to capture the indoor layouts in motion; \textbf{\textit{2)}} incorporating time-series data from inertial measurement units (IMUs) and ambient lighting conditions recorded during data collection to highlight the potential impact of motion speed and lighting on layout estimation accuracy.
It comprises 2.5K sequences, including over 771.3K RGB images and 10 billion event data points. Among these, 39K images are annotated with indoor layouts to facilitate potential research in either event-based or video-based indoor layout estimation. 
Based on the dataset, we propose an event-based layout estimation pipeline with a novel event-temporal distribution feature module to effectively aggregate the spatio-temporal information from events. 
Meanwhile, we introduce a spatio-temporal feature fusion module, which can be easily integrated into the transformer module for fusion purposes. Then, we conduct benchmarking, and extensive experiments on our Ev-Layout dataset show that our approach significantly improves the accuracy of dynamic indoor layout estimation compared to event-based methods.


\end{abstract}

\section{Introduction}\label{sec:introduction}

Indoor layout estimation aims to obtain pixel-level labels that indicating the positions of walls, floors, and ceilings while clearly defining their boundaries. This spatial layout information significantly enhances scene understanding and provides essential support for various visual tasks in indoor environments, including scene reconstruction, autonomous robot navigation, virtual reality, and augmented reality.~\cite{shen2024360, guo2013support}
As a result, several high-quality datasets for indoor layout estimation have been developed. These datasets are mainly categorized into two types: those based on 2D perspective images and those based on panoramic images. Datasets based on 2D perspective images.~\cite{dai2017scannet, song2015sun, zheng2019structured3d} are better suited for scenarios requiring quick estimation of indoor layouts, such as augmented reality and service robotics. In contrast, datasets based on panoramic images~\cite{zou2021manhattan, cruz2021zillow} facilitate comprehensive perception of the 360-degree surrounding environment while inherently encoding contextual information about the scene~\cite{zhang2014panocontext}.

When it comes to hosts with high degrees of freedom, such as virtual reality (VR) headsets and mobile robots, layout estimation for faces significant challenges due to the limited texture information or repetitive patterns often present on ceilings and walls. 
These challenges stem from dynamic scenes and visual artifacts, such as motion blur and adverse lighting conditions, which result from the flexible movements of VR users or the rapid navigation of robots in indoor environments \cite{yu2015lsun, zheng2019structured3d, yang2019dula}. To address these issues, there is a need for \textit{new sensing modalities that can capture visual details with high temporal resolution (to mitigate blur artifacts) and a high dynamic range (to handle difficult lighting conditions). }


Event cameras, inspired by biological vision systems, capture per-pixel intensity changes asynchronously \cite{gallego2022event, posch2011qvga}. These sensors offer unique advantages, including high temporal resolution and a high dynamic range, which pave the way for accurate estimation and tracking of indoor layout even during intense motion. Moreover, their high dynamic range provides a significant advantage over traditional RGB cameras, enabling effective layout estimation in complex lighting conditions, such as underexposed or overexposed environments. In this context, this paper presents \textbf{Ev-Layout}, which, to the best of our knowledge, is the \textbf{first} large-scale event-based multimodal dataset designed specifically for indoor layout estimation and tracking.
 \begin{figure*}[t!]
  \centering
  \includegraphics[width=1.0\textwidth]{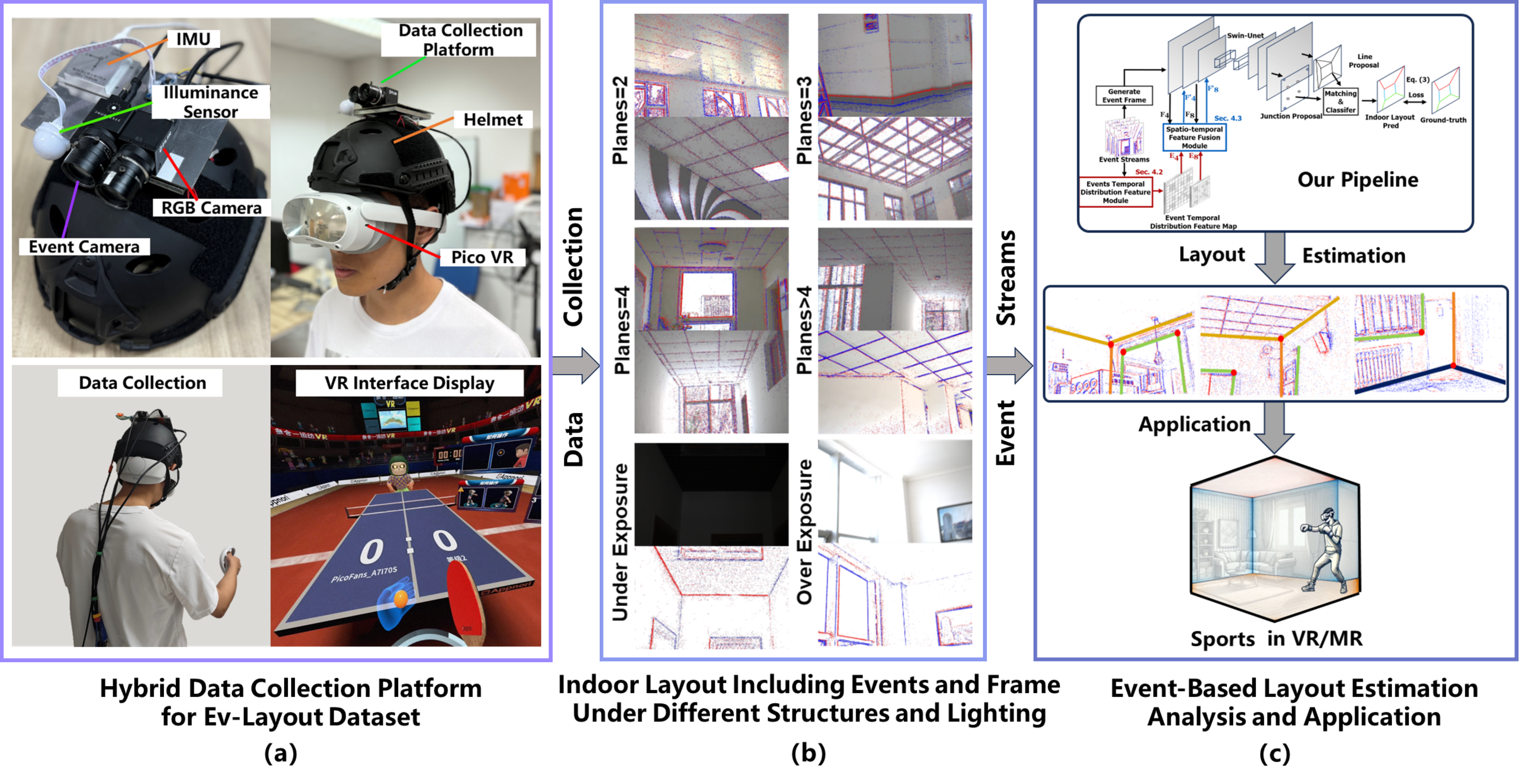}
  \caption{(\textbf{a}) An overview of our data acquisition platform includes an event camera, an RGB camera, an IMU sensor, and a light sensor. When users play sports games with VR/MR devices, the head-mounted data acquisition platform simultaneously captures the indoor layout. 
  (\textbf{b}) Ev-Layout is an event-based dataset for indoor layout estimation under fast motion and complex lighting conditions, potentially supporting spatial computing of event cameras for VR/MR. The dataset is collected via our multi-modal data acquisition platform and includes time-series data with varying lighting intensities, speeds, and structural configurations, accompanied by annotation of continuous layout labels. (\textbf{c}) Based on the dataset, we propose a novel layout estimation pipeline that integrates the temporal distribution feature of events to achieve precise layout estimation, leveraging events' high temporal resolution and high dynamic range. We expect our work could contribute to advancing the application of event cameras in layout estimation for VR/MR.}
  \label{fig:combined}
\end{figure*}

This novel dataset holds several contributions to the community. Firstly, it utilizes a hybrid data collection platform with a head-mounted display and VR interface, see Fig.~\ref{fig:combined}(a). This platform integrates RGB and bio-inspired event cameras to capture indoor layouts in motion. Secondly, it incorporates the time-series data from inertial measurement units (IMUs) and ambient lighting conditions recorded during data collection to highlight the potential impact of motion speed and lighting on layout estimation accuracy, see Fig.~\ref{fig:combined}(b).  
Another highlight is that it includes data under diverse challenging lighting conditions\cite{mohan2022room}. 
Although automatic exposure adjustment by cameras can mitigate some issues caused by challenging lighting, there are inherent limits to sensor sensitivity. According to the statistics of the dataset characteristics (see Fig.~\ref{fig:statistic} in Section~\ref{sec:data_char}), indoor lighting facing windows on a sunny afternoon can reach up to 6K lux, while the lighting in a dim indoor environment on a cloudy afternoon can be as low as 10 to 20 lux. The issues of over-exposure and under-exposure significantly increase the difficulty of indoor layout estimation. Therefore, our dataset includes an illuminance sensor to provide feedback on current environmental brightness and includes extremely over-exposed or under-exposed indoor environment data, as shown in Fig.~\ref{fig:combined}(b). 
As a result, our Ev-Layout dataset comprises 2.5K sequences, including over 771.3K RGB images and 10 billion event data points. Among these, 39K images are annotated with indoor layouts to facilitate potential research in either event-based or video-based indoor layout estimation.



 Based on the dataset, we propose a novel event-based indoor layout estimation pipeline, as depicted in Fig.~\ref{fig:combined}(c) and Fig.~\ref{fig:pipeline_graph}. 
 To exploit the rich motion information contained in events and to accurately estimate indoor layouts under extremely fast motion conditions, we first design a novel event-time distribution feature (\textbf{ETDF}) module to effectively aggregate the spatio-temporal information from events (Sec.~\ref{sec:e_features}). This feature module allows for effective edge information to be obtained for layout estimation even with very sparse event data. Meanwhile, to facilitate ETDF module, we propose a transformer-based spatio-temporal feature fusion (\textbf{SFFM}) module (Sec.~\ref{sec:e_fusion_module}). This module is highly flexible and can be integrated with any transformer module, incorporating the information from the event-time distribution features into the existing feature maps. Finally, we conduct a benchmark based on our Ev-Layout dataset. The results show that our approach significantly improves the accuracy of dynamic indoor layout estimation compared to other event-based methods. In summary, our major contributions are:
\begin{itemize}
 \item We collect a large-scale event-based multi-modal dataset for estimation and tracking indoor layouts with video and events, time-series data from IMUs, and ambient lighting conditions, for indoor layout estimation and tracking.
 \item We propose a novel event-based layout estimation framework with two key modules: 1)  We propose an event-time distribution feature module to obtain effective edge information with an extremely small temporal resolution (less than 1 ms), and 2) a spatio-temporal feature fusion module, which can be flexibly integrated into the transformer for fusion purposes.
 \item We conduct a benchmark, showing the superiority of our Ev-Layout dataset, and proposed framework that significantly improves the accuracy of dynamic indoor layout estimation.
 \end{itemize}

\section{Related Work}
\subsection{Conventional Layout Estimation and Dataset.}
Indoor layout estimation has been explored in various ways. Initially, Hedau et al. \cite{hedau2009} used the Manhattan World assumption \cite{coughlan1999} as a prior for determining room shapes. In the past, researchers used handcrafted features \cite{schwing2012, delpero2012, dijkstra1959}, followed by vanishing point detection and hypothesis generation. Recent methods employ convolutional neural networks (CNNs), such as Mallya and Lazebnik's \cite{mallya2015} use of structured edge detection forests with CNNs to predict edge probability masks, and Lin et al.'s \cite{lin2018} end-to-end CNN for pixel-wise indoor image segmentation. DeepRoom3D \cite{huang2018deep} predicts cuboids using an end-to-end CNN, while RoomNet \cite{lee2017roomnet} directly predicts ordered key points in indoor layouts. Other recent methods combine CNNs with geometric priors to optimize indoor layout estimation \cite{ren2017, zhao2017, yan2020, kruzhilov2020}. Some approaches avoid these geometric constraints. Stekovic et al. \cite{stekovic2020} solve discrete optimization problems over 3D polygons using RGB and depth information. Howard Jenkins et al. \cite{howardjenkins2018} form 3D models from video sequences using plane detection, and Yang et al. \cite{yang2016holistic} use plane, depth, and vertical line detections for general indoor layout estimation. Recently, Gillsjö et al.\cite{gillsjo2022semantic} predict a semantic wireframe based on room geometry using a Graph Convolutional Network, but this does not account for plane instances. Then, Gillsjö et al.\cite{gillsjo2023polygon} introduce a neural network-based method for semantic plane detection using polygon representations, which improves indoor layout estimation using a graph-based approach.
The most popular indoor layout estimation datasets, such as LSUN \cite{yu2015lsun}, HEDAU \cite{hedau2009recovering}, and Structured3D \cite{zheng2019structured3d}, are based on single images and do not include continuous video data. Some datasets aim at room reconstruction, such as the TUM dataset \cite{schubert2018tum}, and ScanNet dataset \cite{dai2017scannet}, which do have continuous video data but lack corresponding continuous labels. The Michigan Indoor Corridor Dataset \cite{huang2018corridor} includes videos and continuous annotated labels, but only has four video sequences and is limited to corridor environments, making the dataset too small in scale. In a recent study, Meta utilized Aria glasses equipped with RGB cameras to collect video datasets from 25 different indoor rooms~\cite{straub24efm}. In contrast, the Ev-Layout dataset not only incorporates event cameras as an additional modality but also captures data across over 200 real-world scenes, offering a larger scale and greater diversity of indoor environments.
\\
\subsection{Event Camera.}
Event cameras are advanced visual sensors that differ from standard RGB cameras. They detect changes in environmental brightness asynchronously, recording visual data as events. An event is denoted as $(x, y, t, p)$, where $x$ and $y$ are the pixel coordinates, $t$ is the time of occurrence, and $p$ indicates event polarity. Unlike RGB cameras that stream images at a certain frequency, event cameras output event points asynchronously. For each pixel $\mathbf{u} = [x, y]^T$, if the captured luminance $L$ changes beyond a contrast threshold $C$ at time $t$, an event point $\mathbf{e} = (x, y, t, p)$ is triggered. The increase and decrease of luminance result in positive $(p = +1)$ and negative $(p = -1)$ polarities, respectively.
\begin{equation}
L = p \left( L(x, y, t) - L(x, y, t) \right) \geq C
\end{equation}
Where $t$ is the time of the last event point generated at that pixel. An event camera captures an event as long as a change is detected, without global synchronization, thus being able to capture high-speed movement with high temporal resolution. Events often occur at the edges, making them advantageous for perceiving indoor layouts. They also feature high temporal resolution, eliminating motion blur from high-speed movements. With a dynamic range of 140 dB, significantly higher than RGB cameras, they perform well under challenging lighting conditions.
Event cameras have been applied in many fields. Event-based methods often start by converting event streams into image-like representations \cite{Maqueda2018,Benosman2013,Zhu2019,Gehrig2019}. These are used in gait recognition \cite{wang2021event, 8953966}, eye tracking\cite{zhao2024rgbe, zhang2024swift}, events-to-video conversion\cite{qu2024e2hqv} and Visible Light Communication\cite{wang2024towards}.
\\
\subsection{Event-based Layout Estimation and Dataset.}
\textit{To our knowledge, there is no direct research on indoor layout estimation using event cameras}. Only a few preliminary works~\cite{munir2022ldnet, reverter2019event, tschopp2021hough} have been done for line segment detection, which is an upstream task to indoor layout estimation, utilizing event cameras for this purpose. However, the task of line detection is less challenging than indoor layout estimation. The earliest event-based line segment methods were applied to high-speed train positioning ~\cite{seifozzakerini2016event}. Subsequently, ELiSeD ~\cite{reverter2019event} utilized a constant velocity model for line segment detection and tracking but struggled with rotating and rapidly moving segments. To enhance robustness, a method based on iterative weighted least squares fitting was proposed~\cite{reverter2019event}. Most of these methods were validated only on simple data, lacking quantitative evaluation and comparison. Later, FE-LSD~\cite{yu2023detecting} collected 800 real event sequences, with line segments annotated using images calibrated with events. This work to some extent alleviates the problem of lack of data.
Our investigation reveals that there are no publicly available large-scale real event datasets for indoor layout estimation benchmarks, which has hindered the development of this field. Some works, such as YUM-vie~\cite{klenk2021tum}, use event cameras to collect indoor data, but they lack continuous layout annotations and scene diversity. \textit{To address this, we propose the \textbf{first} event-based indoor layout estimation dataset, Ev-Layout, aiming to promote the development of event cameras in indoor layout estimation.}

\section{The Ev-Layout Dataset}

\subsection{Dataset Modalities}
We build a data collection platform for dataset collection. It incorporates four different sensory modalities:  1) images captured by an RGB camera, 2) event streams captured by an event camera, 3) motion time-series data captured by an IMU sensor, and 4) brightness in the shooting direction measured by an illuminance sensor.

\textbf{\textit{Event Streams:}} The event streams are collected by an Prophesee EVK4 event camera\cite{prophesee2024evk4} with a resolution of 1280x720. As the data collection platform moves, the event camera records indoor information in response to changes in intensity. The high temporal resolution and high dynamic range of event cameras enable capturing the clear edges of the indoor scenes even under high-speed motion and challenging lighting conditions.

\textbf{\textit{RGB Images:}} The RGB images are collected by the FLIR RGB camera (FLIR BFS-U3 16S2C)\cite{flir_blackfly}, with a resolution of 1440x1080, recording the videos of different indoor scenes.

\textbf{\textit{IMU Time-series Data:}} The IMU time-series data are collected by the inertial measurement unit (WHEELTEC N100)\cite{ros2_wheeltec_n100_imu}, with a sampling rate of 400 Hz. The IMU sensor is placed behind the FLIR and EVK4 cameras to record the 3-axis acceleration and 3-axis angular rate of the movement of the data collection platform.

\textbf{\textit{Illuminance Sensor:}} The illuminance sensor measures light intensity in lux, with a sensing range of 0-65535 lux. It is aligned with the camera's shooting direction to perceive the environmental brightness during data collection.

 \begin{figure}[t!]
  \begin{center}
    \includegraphics[width=\columnwidth]{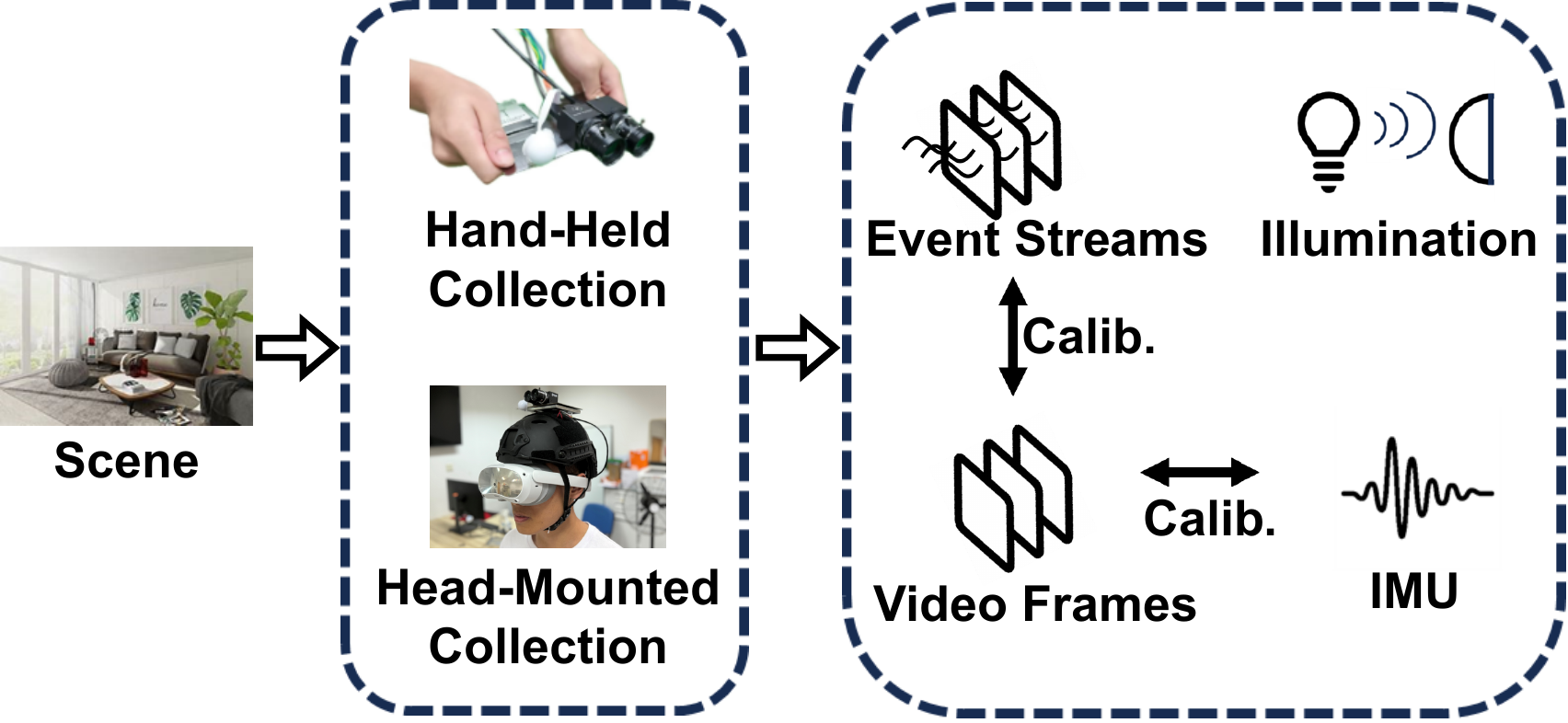}
  \end{center}
  \caption{Data collection, processing, and annotation process.}
  \label{fig:flow_chart}
\end{figure}

\subsection{ Dataset Curation }
\label{sec:Data_interpretation}
\noindent \textbf{Data Collection Setup. }
As shown in Fig.~\ref{fig:combined}(a), the data acquisition equipment consists of an event camera and an RGB camera, with a baseline set at 3 cm. These cameras form a stereo vision system through vision calibration. Due to the different resolutions of the event camera and RGB camera,  images are rescaled to the resolution of the event camera, i.e.,  1280x720. The frame rate of the RGB camera is set at 100 Hz to ensure clear images are captured even when the data collection platform is in rapid motion. This setup is intended to enable more precise data annotation on the RGB images.

Besides, the platform is equipped with an inertial sensor to profile the movement of the entire platform, and an illuminance sensor to measure the environmental brightness when collecting data. Throughout the data collection process, the exposure and focal length of both the event camera and RGB cameras are fixed. This controlled setup ensures that the RGB camera can capture underexposed and overexposed data by keeping the exposure time and aperture variables constant, thereby maintaining a strong correlation between the brightness information captured by the RGB camera and the illuminance sensor.

\begin{figure}[t!]  
  \begin{center}
    \includegraphics[width=\columnwidth]{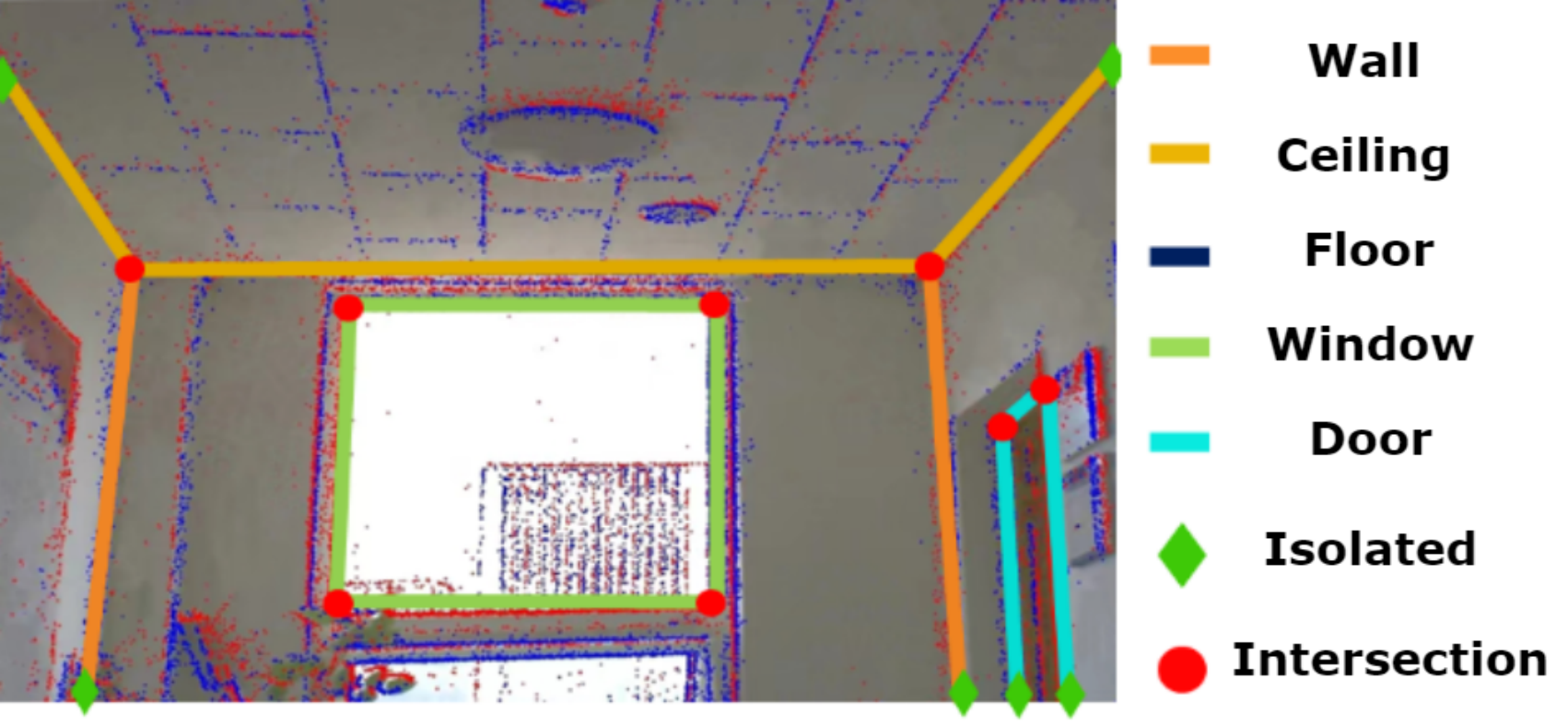}
  \end{center}
  \caption{Visualization of data annotation.}
  \label{fig:label} 
\end{figure}

Fig.~\ref{fig:flow_chart} presents the data collection process. For each scene, the handheld subset was collected by aiming the lens towards the targeted scene and shaking the platform for 3 seconds. 
The collection platform was calibrated to ensure time synchronization across multiple devices. The obtained events and video frames are spatially aligned to ensure pixel-level correspondence \cite{liu2024lecalib}. The IMU and RGB cameras are also aligned to ensure they operate in the same coordinate. Finally, the video frames are used for annotating the indoor layout.

\noindent \textbf{Acquisition Protocol.}
The entire dataset comprises two major subsets: 1) a large subset collected in handheld mode and 2) a small subset collected in head-mounted mode. 


\textbf{\textit{Collection in the handheld mode:}} To improve data acquisition efficiency, most of the data for Ev-Layout was collected in handheld mode, as show in Fig.~\ref{fig:flow_chart}. During the collection process, the handheld platform simulates head movements by shaking the hands to align the platform with the target layout, covering multiple angles.

\textbf{\textit{Collection in the head-mounted mode:}} As shown in Fig.~\ref{fig:combined}(a), the entire data collection platform was fixed to the top of a helmet, and five participants wore a PicoVR device~\cite{pico4pro} along with our acquisition equipment.
It has been shown that the maximum angular velocity of human head movement can reach up to 12.8 rad/s~\cite{miller2020envelope}. As a result, at such high angular speeds, video-based layout estimation methods are severely affected by motion blur. In contrast, event-based layout estimation methods remain unaffected. To obtain realistic head movement data while using VR devices, the data collection for our Ev-Layout dataset covers a range of angular velocities from 0 to 12 rad/s. In addition to the common VR actions such as head tilting and turning, we also capture indoor layout data under more intense head movements. This ensures that our dataset covers the scenarios involving rapid head motions, providing comprehensive data for layout estimation even in challenging conditions. We asked the participants to play the ``All-In-One Sports VR'' game (see Fig.~\ref{fig:combined}(a)), which includes boxing, basketball, table tennis, tennis, cricket, and badminton, on the Pico while collecting the subset.
The head movements are categorized into four levels based on intensity: very slowly (looking around left and right, raising head up and down), gentle (cricket, basketball), moderate (badminton, tennis), and intense (boxing, table tennis). The acquisition device was activated during the VR sports activities, and the 5-second segment with the most intensive head movement was selected. Moreover, five different representative indoor environments, including a gym, leisure area, office, corridor, and meeting room, were chosen for the data collection.

\begin{table}[t!]
\centering
\resizebox{0.70\linewidth}{!}{
\begin{tabular}{lll}
\hline
Label        & Counts & Occurrence \\ \hline
Ceiling-Wall & 55432  & 26.31\%    \\
Wall-Wall    & 49007  & 23.26\%    \\
Floor-Wall   & 25600  & 12.15\%    \\
Window-Wall  & 36589  & 17.37\%    \\
Door-Wall    & 43996  & 20.88\%    \\ \hline
Total        & 210642 & 100.00\%   \\ \hline
\end{tabular}}
\vspace{30pt}
\caption{Label counts and their occurrences.}
\label{table:label_counts}
\end{table}

\noindent \textbf{Data Annotation.}
The data collection results in a large-scale dataset containing 771,300 images captured by the RGB camera. We selected 39,200 RGB images for annotation using the LabelMe 
software\cite{labelme}. For each image, we annotate the coordinates of endpoints and the line segments formed by these endpoints. The endpoints are categorized into two types: the intersection points of two or more line segments, and the isolated endpoints that exist on only a single straight line. The line segments are divided into five categories: wall-wall intersection (wall), wall-floor intersection (floor), wall-ceiling intersection (ceiling), wall-window intersection (window), and wall-door intersection (door). Table~\ref{table:label_counts} shows occurrences of the five categories. Fig.~\ref{fig:label} presents an example of such annotations on a single image with the corresponding event stream recorded at the same time.

\begin{figure}[t!]
  \begin{center}
    \includegraphics[width=\columnwidth]{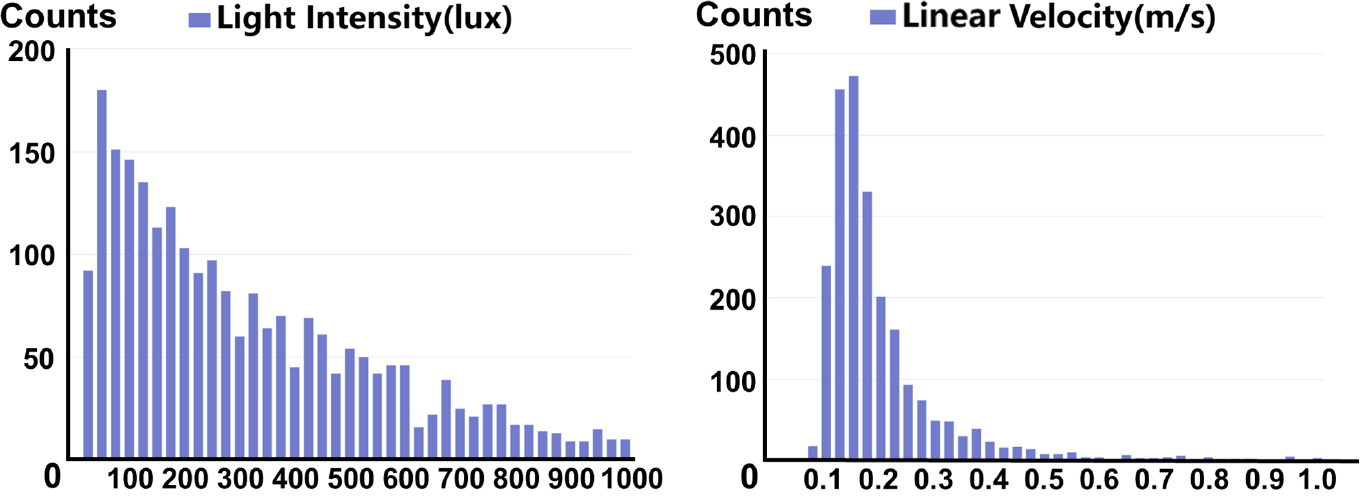}
  \end{center}
  \caption{Statistical histogram depicting the ambient light intensity (left) and movement speed (right).}
\label{fig:statistic}
\end{figure}

\subsection{ Data Characteristics}
\label{sec:data_char}
Our dataset includes the multimodal data collected from an event camera, an RGB camera, IMU sensors, and an environmental illuminance sensor. The RGB camera provided 771,300 RGB images and the event camera generated event streams with over 10 billion event points. The entire dataset comprises 2,500 sequences, covering 200 different scenes.
As the movement is the most distinctive characteristic of our dataset from other existing ones. 
We first present the distribution of the motion intensity of the recorded sequences by calculating the moving speed of the data collection platform from the IMU time-series data over a 3-second interval. 

For data collected by handheld methods, the frequencies histogram over different linear velocities is calculated. As shown in Fig.~\ref{fig:statistic}(right), it is evident that the primary collection speed of the dataset is 0.15 m/s. Also, there are sequences with average collection speeds higher than 0.8 m/s and lower than 0.1 m/s, which better reflect the influence of movement on layout estimation. 

\begin{figure}[t!]
  \begin{center}
    \includegraphics[width=\columnwidth]{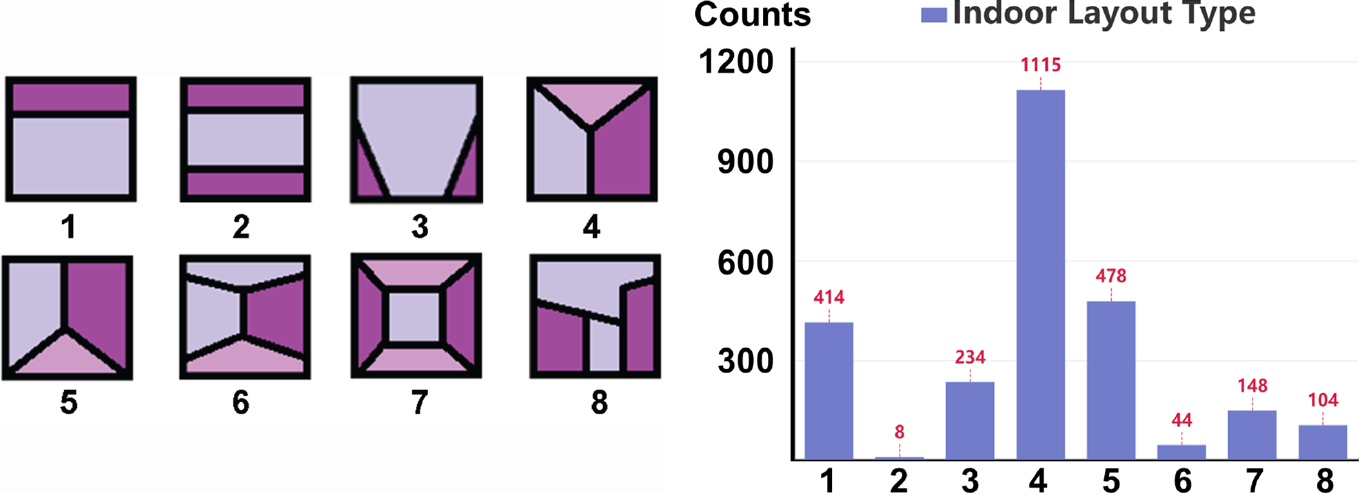}
  \end{center}
  \caption{ Statistics of eight types of indoor layout(left) and different types of indoor layout(right).}
\label{fig:category}
\end{figure}

\begin{figure}[t!]
    \begin{center}
      \includegraphics[width=\columnwidth]{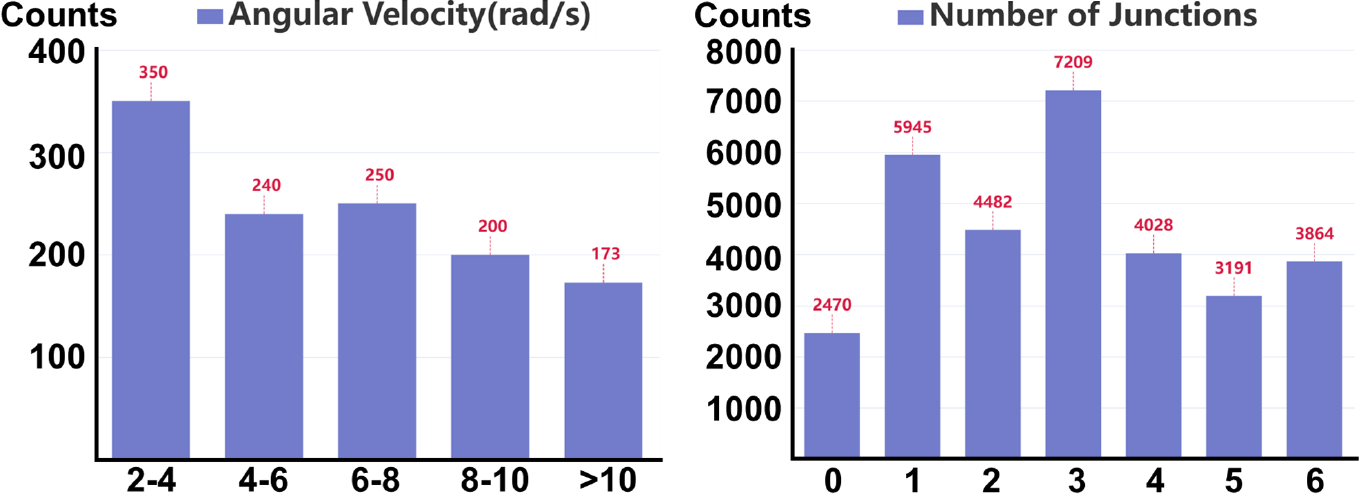}
    \end{center}
    \caption{Statistics of statistical histogram depicting the angular velocity(left) and different junction numbers(right).}
\label{fig:statistic2}
\end{figure}

\begin{figure*}[t!]
  \centering
  \includegraphics[width=1.0\textwidth]{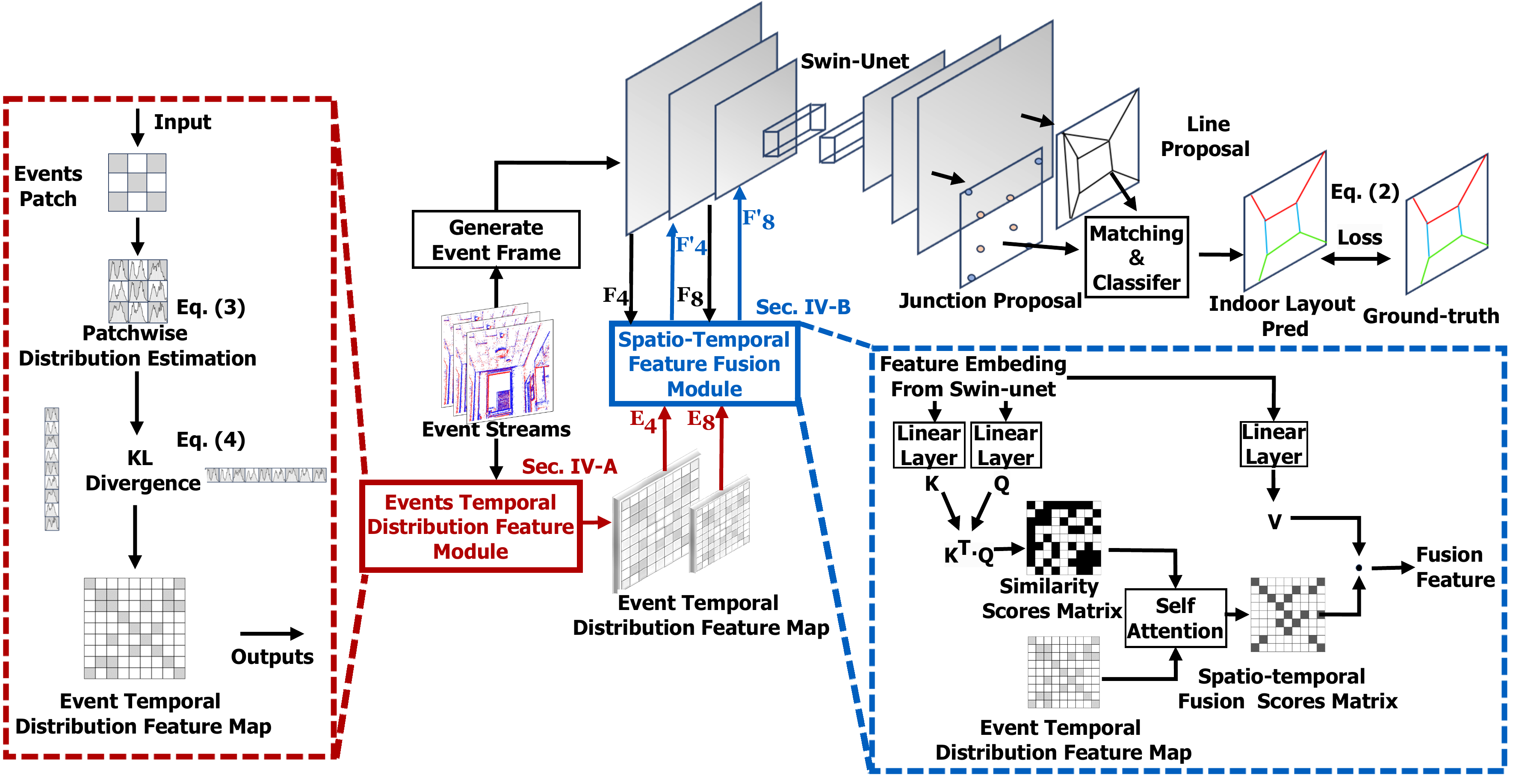}
  \caption{An overview of the proposed benchmarking approach of Ev-Layout.}
  \label{fig:pipeline_graph}
  \vspace{-10pt}
\end{figure*}

For data collected using a head-mounted display method, it is more appropriate to classify the collected data based on angular velocity, as head movements primarily consist of rotational motions. As shown in Fig.~\ref{fig:statistic2}(left), the angular velocity of the collected data is mainly distributed between 4 and 8 rad/s, with the maximum angular velocity of head movements reaching 12.8 rad/s~\cite{miller2020envelope}. Therefore, the dataset also includes data with angular velocities ranging from 10 to 12 rad/s.
We then show the dynamics of lighting conditions of the scenes in our dataset by plotting the frequencies histogram of the light intensity (lux) for all sequences. From Fig.~\ref{fig:statistic}(left), it can be observed that the primary illuminance range of the dataset is between 100-400 lux. This conforms to the typical range of indoor room brightness~\cite{tiller1995perceived}.  Moreover, there are also sequences with extreme lighting conditions, with illuminance levels below 50 lux and above 800 lux.

To fully showcase the diversity of the Ev-Layout dataset, we provide a statistical analysis of the dataset from multiple perspectives. We find that the complexity of indoor scenes is closely related to the number of junction points in the images. Therefore, the dataset is first categorized based on the number of junction points in the images, ranging from 0 to over 5, as depicted in Fig.\ref{fig:statistic2}(right). The y-axis represents the number of images corresponding to each category. This frequency histogram demonstrates that the Ev-Layout dataset exhibits a relatively balanced distribution in terms of the number of junction points, i.e., the complexity of the scenes.

Moreover, we collect eight types of indoor layouts following the setting in RoomNet~\cite{lee2017roomnet}, as shown in Fig.~\ref{fig:category}. In this categorization, the eighth type represents complex scenes that are difficult to classify into the first seven types. We then show the distribution of the types of all the images in the dataset in Fig.~\ref{fig:category}. It can be seen that the Ev-Layout dataset has a higher number of images in categories 1, 3, 4, 5, 7, and 8, covering common indoor layout scenes as well as some complex scenes.

\section{The Proposed Pipeline}
\label{sec:feature}

\noindent\textbf{Pipeline Overview.}
\label{sec:pipeline}
An overview of the proposed pipeline is shown in Fig.~\ref{fig:pipeline_graph}. Initially, pure event data is processed into two forms: a 2D event image and an Event Temporal Distribution Feature Map. These two representations contain different types of spatio-temporal information. The 2D event image ignores the temporal information of event triggers, focusing only on the spatial information. In contrast, the feature map is derived solely from the temporal distribution characteristics of events, disregarding the spatial information of events. The two features can complement each other. After dividing the image into many patches, the Event Temporal Distribution Feature (\textbf{ETDF}) Module (Sec.~\ref{sec:e_features}) approaches the similarity between patches by first obtaining the probability distribution function corresponding to each patch. Then, the similarity between patches can be quantified using KL divergence, ultimately generating the ETDF. The Spatio-temporal Feature Fusion Module (\textbf{SFFM}) approaches the computation of the similarity matrix from a transformer perspective, replacing the matrix similarity with the similarity of the probability distribution function (Sec.~\ref{sec:e_fusion_module}). This allows the Event Temporal Distribution Feature Map to be integrated with the transformer block. 

The encoder part of the network uses Swin-Unet\cite{cao2022swin}, generating two versions of the Event Temporal Distribution Feature Map with different patch sizes(4, 8) and integrating them with Swin-Unet at two different stages. This operation allows information at different scales to be fused within the network. The decoder part employs a structure similar to HAWP\cite{xue2020holistically}. The output of the decoder consists of line proposals and junction proposals. The endpoints of each line segment are matched with their closest junction point, and the matched lines are classified through two fully connected layers. The final loss comprises of losses from the lines, endpoints, and classification.

\begin{equation}
L = L_{\text{line}} + L_{\text{junc}} + L_{\text{cls}}
\end{equation}

\subsection{Events Temporal Distribution Feature Module}
\label{sec:e_features}

\begin{figure}[t!]
  \begin{center}
    \includegraphics[width=\columnwidth]{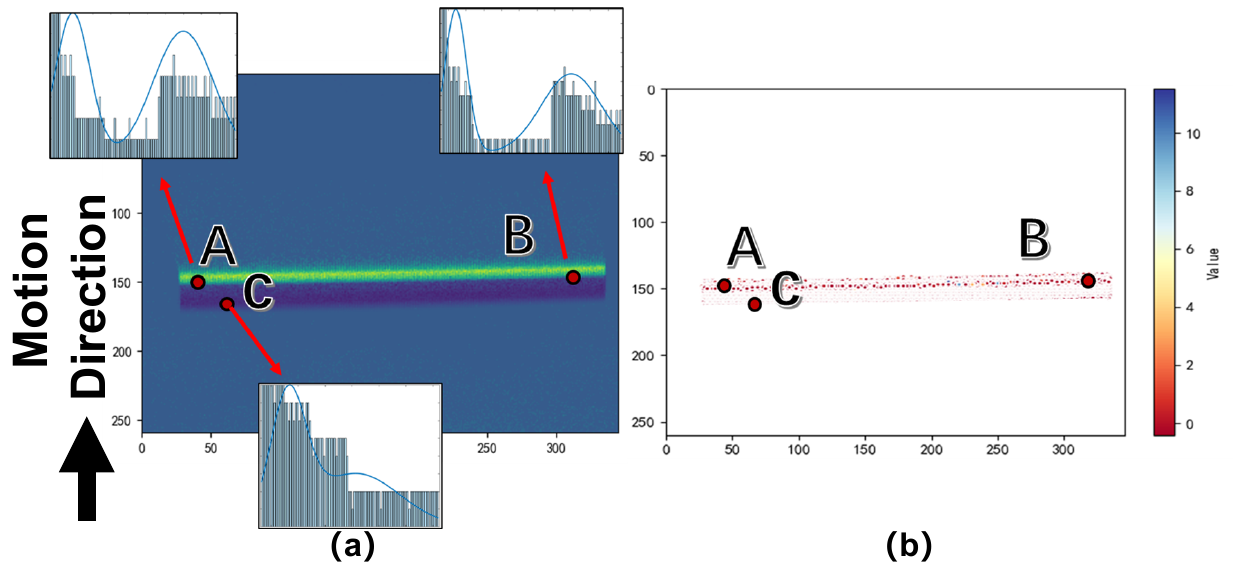}
  \end{center}
  \caption{Illustration of Event Temporal Distribution Features.}
  \label{fig:feature_explain}
\end{figure}

Let's consider the following scenario: an event camera sweeps across a line, recording an event stream. Based on the principles of the event camera, each pixel independently detects changes in the logarithm of light intensity. As shown in Fig.~\ref{fig:feature_explain} for a specific pixel $A$, we collect all events generated at pixel $A$'s location within a time interval \(dt\) (\(dt\)=2ms) and plot a histogram of the frequency distribution with a bin width of 0.2$ms$. The horizontal axis of the histogram represents time $t$, and the vertical axis represents the number of events generated in each bin. If we plot the frequency distribution histograms of pixels $A$, $B$, and $C$ within the same  \(dt\), we observe that the histograms of A and B are significantly more similar to each other than to $C$. This phenomenon can be explained by the fact that, within the  \(dt\) interval, $A$ and $B$ receive two high-frequency signals triggered by the edges of the line at similar times, whereas $C$ only receives the high-frequency signal from the lower edge of the line during the same interval. Thus, the frequency distribution histograms of $A$ and $B$ are more alike. Therefore, we can hypothesize that, within a short time interval, the frequency histograms of pixels located on the same line will exhibit higher similarity.

To verify this hypothesis, we first need to estimate the probability distribution function \( P_{xy}(t) \) from the frequency distribution histogram of each pixel position. Each pixel position will correspond to a probability distribution function. Then, we can use the Kullback-Leibler (KL) divergence\cite{hershey2007approximating} to measure the similarity between the probability distribution functions of different pixel coordinates.

The occurrence of events at pixel coordinates over time is a random process, typically described as a counting process\cite{gu2021spatio}, which tracks the activated number of events. We can model the number of events generated at pixel coordinates over a time interval as a Poisson process. In a time interval of length \( t \), the number of events follows a Poisson distribution with a mean of \( \lambda t \). When considering the characteristics of events further, we distinguish between Homogeneous Poisson processes\cite{pasupathy2010generating} and Inhomogeneous Poisson processes\cite{rathbun1994asymptotic}. Since the event generation rate varies over time, modeling with an Inhomogeneous Poisson process will be more accurate, which can be described as follows:
\begin{equation}
\mathbb{P}_{xy} \{N(t+s) - N(s) = n\} = \frac{e^{-\lambda(t)} [\lambda(t)]^n}{n!}
\end{equation}

Where ${P} \{N(t+s) - N(s) = n\}$ is the probability of \( n \) events occurring in the interval \( t \), and \( \lambda(t) \) is the rate parameter of the Poisson process, which may vary over time in the inhomogeneous case. Thus, for an Inhomogeneous Poisson process, the rate parameter \( \lambda(t) \) becomes a function of time. The formula indicates that the probability of generating \( n \) events at the \( (x,y) \) position within the time interval from \( s \) to \( t+s \) is \( P \). 
Subsequently, the inhomogeneous Poisson process can be used to fit the frequency distribution histogram corresponding to each pixel. Then, the correlation between different pixel points can be estimated using the Kullback-Leibler (KL) divergence. Given that the probability distribution at point \( (x,y) \) is \( P_{xy} \) and the probability distribution at point \( (m,n) \) is \( Q_{mn} \), which can be formulated as: 
\begin{equation}
\mathbb{D}_{KL}(P_{xy} \| Q_{mn}) = \sum_{i} P_{xy}(i) \log \frac{P_{xy}(i)}{Q_{mn}(i)}
\end{equation}
We can quantify the similarity between the probability distributions of different pixel locations.
A smaller KL divergence \( D \) indicates a smaller gap between the probability distribution functions, implying that the distributions at \( (x,y) \) and \( (m,n) \) are more likely to lie on the same edge. 

By calculating the probability distribution functions for all events in the previously mentioned scenario, and identifying all points with a KL divergence less than a given threshold from the probability distribution function of point \( A \), we can mark these points in red in Fig.\ref{fig:feature_explain}(b). Darker shades of red on the points indicate smaller differences. This visualization supports our hypothesis that pixels located at the same edge exhibit lower KL divergence and higher correlation. We designate this method of measuring correlation as the Event Time Distribution Feature.

\begin{figure}[t!]
  \begin{center}
    \includegraphics[width=\columnwidth]{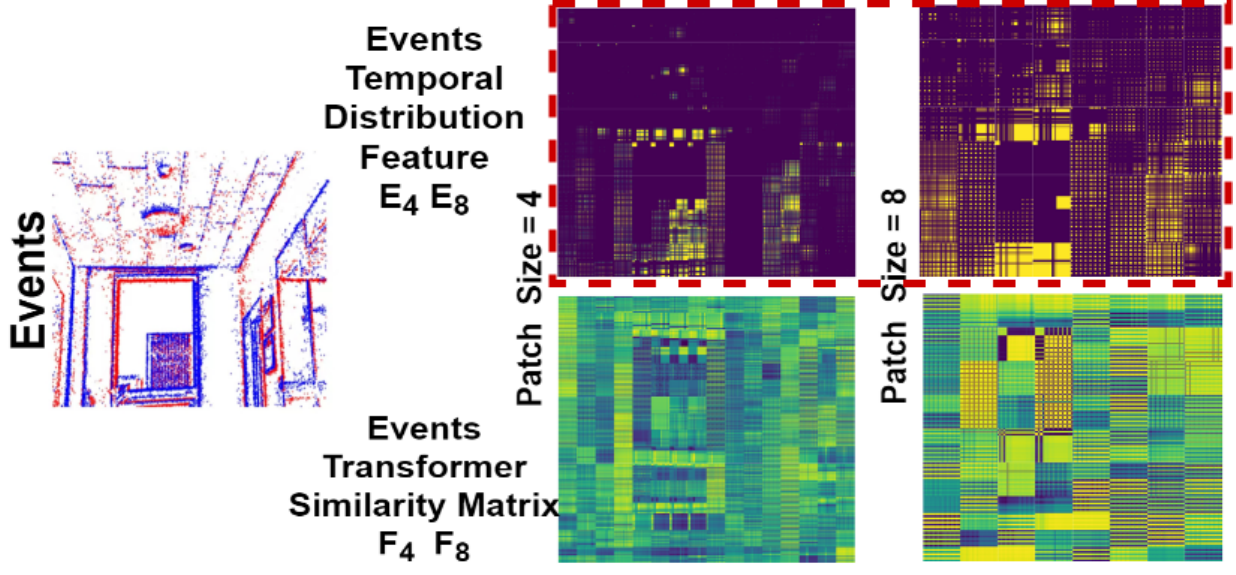}
  \end{center}
  \caption{Visualization of feature maps.}
  \label{fig:feature_visual}
\end{figure}

\subsection{Spatio-temporal Feature Fusion Module}
\label{sec:e_fusion_module}
Estimating the probability distribution function for each pixel and calculating the similarity between pixels results in a computationally intensive process. To integrate Event Temporal Distribution Features with mainstream network architectures, we drew inspiration from the Vision Transformer\cite{dosovitskiy2020image}, which divides an image into patches. Similarly, a \(224 \times 224\) image is divided into several \(16 \times 16\) patches. Each patch contains 256 pixels, with each pixel corresponding to a distribution function based on the conclusions drawn in Sec.~\ref{sec:e_features}. The sum of several independent distribution functions is still a distribution function\cite{barakat1976sums}, so we combine all 256 pixels in one patch to obtain a single probability distribution function. Thus, we have 196 patches corresponding to different probability distribution functions.

\begin{table*}[t]
    \centering
    \resizebox{\textwidth}{!}{%
    \scalebox{1.}{
    \begin{tabular}{cccccccccc} 
        \toprule
        \textbf{Methods} & \textbf{sAP$^5$} & \textbf{sAP$^{10}$} & \textbf{sAP$^{15}$} & \textbf{sAP$^m$} & \textbf{jAP$^{0.5}$} & \textbf{jAP$^{1.0}$} & \textbf{jAP$^{2.0}$} & \textbf{jAP$^m$} \\
        \midrule
        FE-LSD+Classifier Layer(Voxel Grid) & 22 & 25 & 25.7 & 24.23 & 10.5 & 27.6 & 45 & 27.1 \\
        FE-LSD+Classifier Layer(EST) & 27.8 & 32.8 & 34.2 & 31.6 & 10.5 & 23.5 & 38.9 & 24.3 \\
        FE-LSD+Classifier Layer(EC+SAE) & 31.0 & 35.0 & 36.1 & 34.0 & 13.4 & 28.3 & 44.6 & 28.7 \\
        FE-LSD+Classifier Layer(EC+SAE*) & 35.3 & 40.1 & 41.4 & 38.9 & 13.1 & 27.6 & 43.9 & 28.2 \\
        \midrule
        Ours(Voxel Grid) &  \textbf{46.0} &  \textbf{49.5} &  \textbf{50.3} & \textbf{48.6} & 14.7 & 32.9 & 49.4 & 32.3 \\
        Ours(EST) & 45.2 & 48.3 & 49.1 & 47.5 & 16.7 & 34.2 & \textbf{49.7} & 33.5 \\
        Ours(EC+SAE) & 44.5 & 47.4 & 48.4 & 46.7 & 16.4 & 34.0 & 49.0 & 33.1 \\
        Ours(EC+SAE*) & 43.2 & 46.1 & 46.7 & 45.3 &  \textbf{17.0} &  \textbf{34.3} & 49.5 & \textbf{33.6} \\
        Ours without ETDF & 43.3 & 46.7 & 47.6 & 45.8 & 13.6 & 30.6 &  46.2 & 30.1 \\
        
        \bottomrule
    \end{tabular}
    }
    }
    \vspace{1pt}
    \caption{Evaluation results for indoor layout estimation.}
    \label{tab:your_table_label}
\end{table*}

Considering the \( K^T \cdot Q \) operation in self-attention, which essentially uses matrix multiplication to compute the correlation between image patches, resulting in a \( 196 \times 196 \) similarity scores matrix, we replace the matrix multiplication with KL divergence to measure the correlation between different patch probability distribution functions. This yields a similarity scores matrix of the same \( 196 \times 196 \) size, which we define as the Event Time Distribution Feature Map. The specific computational process is shown in Fig.~\ref{fig:pipeline_graph}, highlighted in blue.

As shown in Fig.~\ref{fig:pipeline_graph} in the spatio-temporal feature fusion module, $E_{4}$ and $E_{8}$ are outputs from the events temporal distribution feature module and $F_{4}$ and $F_{8}$ represent the feature embeddings from swin-unet. The numbers 4 and 8 indicate different patch sizes. After passing through the spatio-temporal feature fusion module, the fused features $F'_{4}$ and $F'_{8}$ are obtained, which are then finally input into swin-unet. We visualize $E_{4}$, $E_{8}$, $F_{4}$, and $F_{8}$. As shown in Fig.~\ref{fig:feature_visual}, it can be seen that features from both temporal and spatial jointly reflect the structural information of the room.

The Event Temporal Distribution Feature Map has the same dimensions as the similarity scores matrix obtained by the transformer block. By fusing the two matrices with an attention layer, we can integrate the Event Time Distribution Feature with any transformer architecture.transformer block.

\section{Experiment and Evaluation}
\label{sec:evaluation}

\noindent \textbf{Implementation and Evaluation Metrics.}
We split the handheld subset into the training set and test set with an 8:2 ratio. 
We utilize the sAP (\textit{Structural Average Precision}) metric to evaluate line prediction accuracy and the jAP (\textit{Junction Average Precision}) metric for junction prediction accuracy.

sAP (structural average precision)\cite{zhou2019end} is the area under the precision and recall curve for a set of scored detected line segments. Let $\hat{L} = \left( \hat{p}_1 \; \hat{p}_2 \right) \in \mathcal{L}_g$ be any line segment in the set of true line segments. A detected line segment $L = (p_1 \; p_2)$ is a true positive if
\begin{equation}
\min_{\hat{L} \in \mathcal{L}_g} \delta(L, \hat{L}) = \|p_1 - \hat{p}_1\|_2^2 + \|p_2 - \hat{p}_2\|_2^2 \leq \beta,
\end{equation}
where $\beta$ is a design parameter. We also require that only one detection is matched to each true line segment. Any extra predictions are marked as false positives. In our experiments, we evaluate the metric at $\beta = 5, 10, 15$ at $64 \times 64$ resolution and denote the metrics sAP\textsuperscript{5}, sAP\textsuperscript{10} and sAP\textsuperscript{15} respectively. We calculate sAP for each separate label and then take the mean to form msAP. We also calculate $sAP^m = \sum_\beta sAP^\beta / 3$ for each label and take its mean across labels as sAP\textsuperscript{m}.

jAP (junction average precision) is analogous to sAP. Instead of the criteria in Equation (5) we take the Euclidean distance between the junction and the closest ground truth junction of the same label. The thresholds are 0.5, 1.0 and 2.0 and jAP\textsuperscript{m}, jAP\textsuperscript{m} follows as before.

\begin{figure*}[t!]
    \centering
    \begin{minipage}{1.\textwidth}
        \centering
        \includegraphics[width=\textwidth]{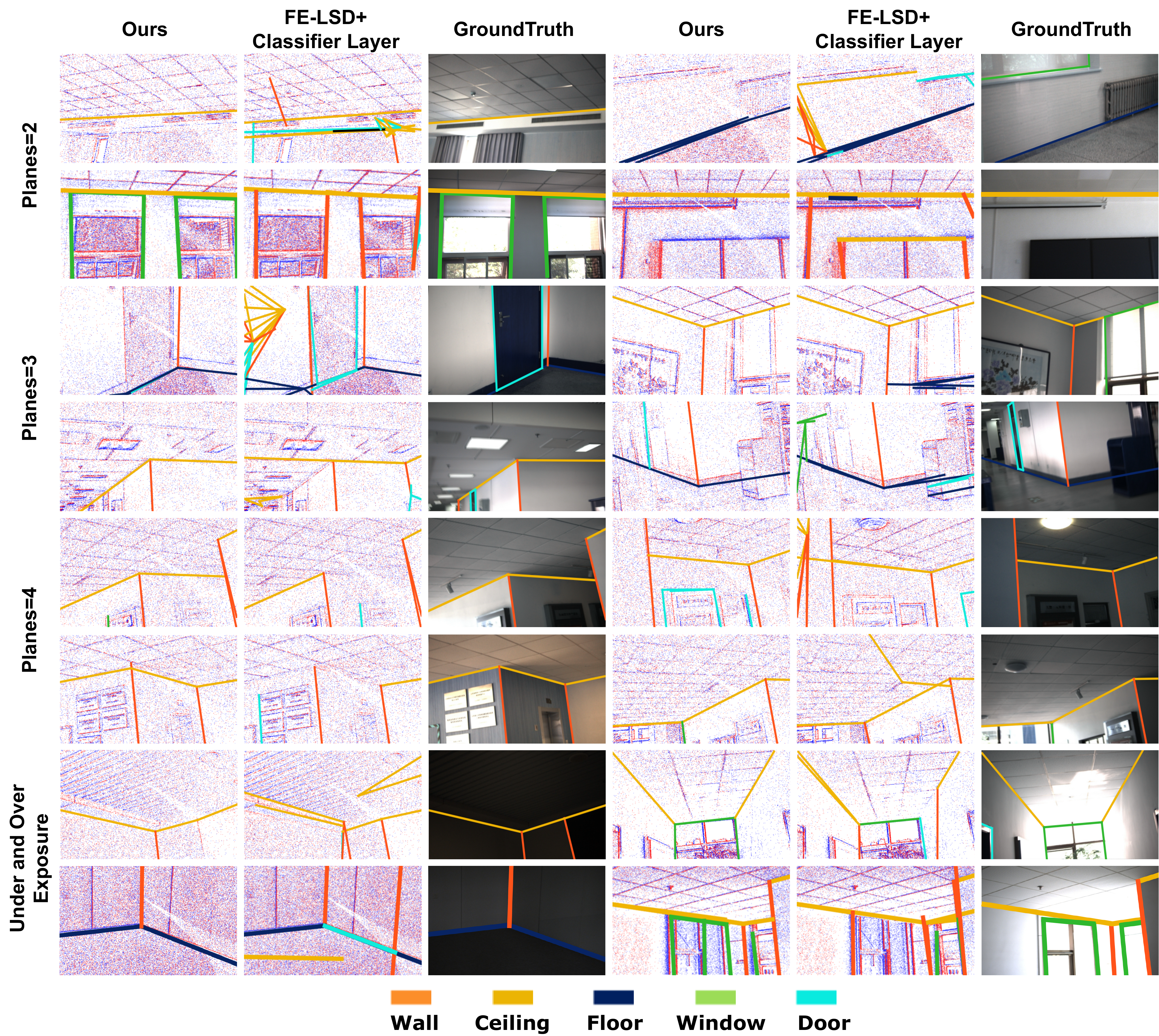}
        \caption{Examples of layout estimation results from different methods for the scenes with different number of planes and abnormal exposure.}
        \label{fig:image3}
    \end{minipage}
\end{figure*}

\begin{figure*}[htbp]
    \centering
    \includegraphics[width=\textwidth]{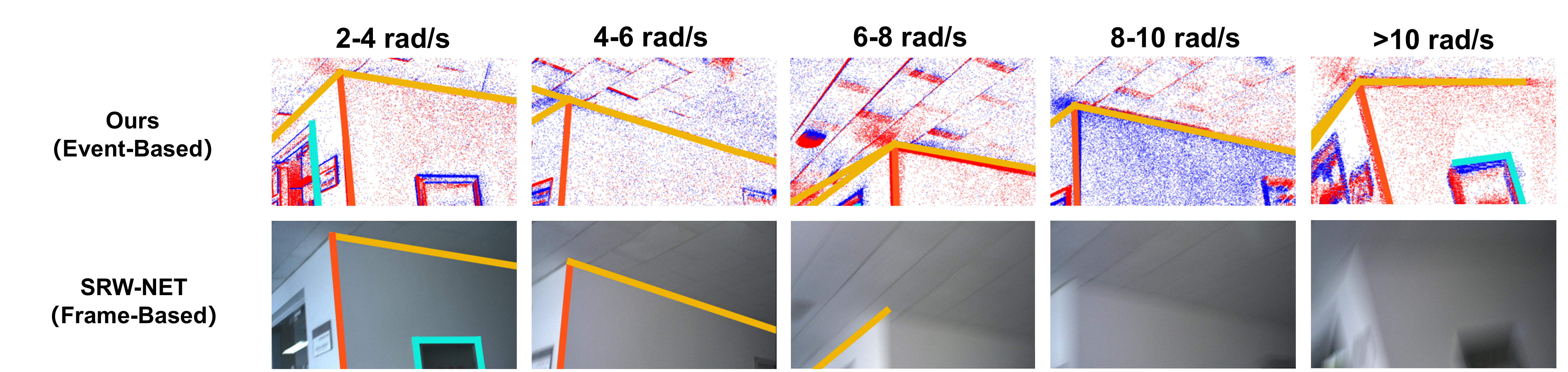}
    \caption{Comparison of visualization results between event-based method (our proposed approach) and frame-based method (SRW-NET) at different angular velocities (rad/s).}
    \label{fig:diff_angular_v}
\end{figure*}

\subsection{Evaluation results} 

Due to the lack of event-based indoor layout estimation methods, we implemented the line detection framework from FE-LSD~\cite{yu2023detecting} and added two additional fully connected layers to classify the proposal lines and junctions into the indoor layout. FE-LSD itself contains both event and frame branches; we remove its frame branch to form an event-based method. We also implement our proposed approach and its variant method without ETDF (Events Temporal Distribution Feature) module for comparison. All the event-based methods take the VoxelGrid representation~\cite{Zhu2019} of each 10ms event stream as inputs. All methods were trained on the part of the handheld subset, and tested on the part of the handheld(2561) and the whole head-mounted subset(2380). The evaluation results are shown in Table~\ref{tab:your_table_label}.

By comparing results among the event-based methods, we observed that our proposed approach
outperforms the competing methods significantly for all the evaluation metrics. Meanwhile, the comparison between our proposed approach with or without ETDF demonstrates the effectiveness of our design of events' temporal distribution features. 

\begin{figure}[t!]
  \begin{center}
    \includegraphics[width=\columnwidth]{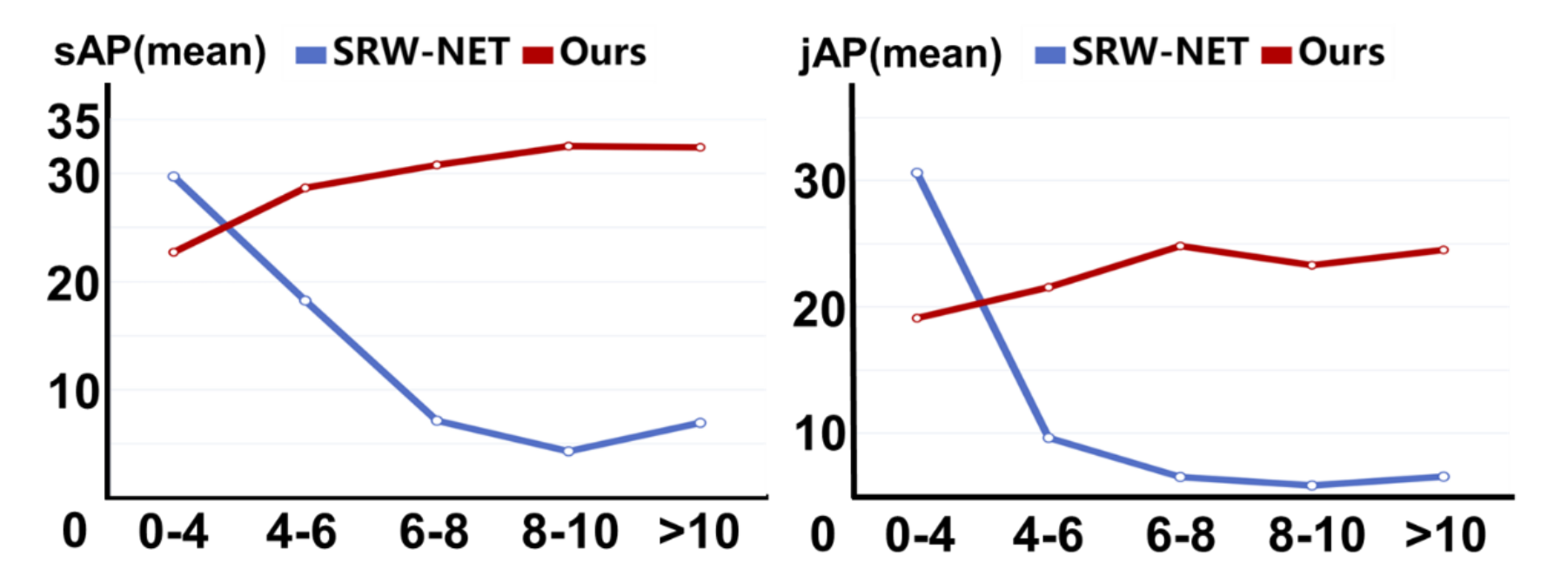}
  \end{center}
  \caption{Accuracy comparison at different angular velocities (rad/s).}
\label{fig:diff_speed}
\end{figure}

Considering that various event-stream representations may have different impacts on the accuracy of layout estimation, we implement and evaluate four state-of-the-art representations with the backbone of our proposed approach on our dataset, including VoxelGrid~\cite{Zhu2019}, Event Spike Tensor (EST)~\cite{Gehrig2019}, Counts of Events plus Surface of Active Events (EC+SAE), and Counts of Events plus Surface of Active Events without polarity (EC+SAE*)~\cite{Benosman2013, Maqueda2018}. Table~\ref{tab:your_table_label} presents the experimental results of our proposed approach and FE-LSD with the four representations. It can be seen that for our proposed approach, the highest $sAP^{m}$ is achieved when event streams are represented as VoxelGrid, and the highest $jAP^{m}$ is achieved when represented as EC+SAE*. Additionally, with the backbone of our proposed approach, all four representations achieve higher $sAP^{m}$ and $jAP^{m}$ scores than those with FE-LSD.

\begin{figure}[H]
    \begin{center}
    \includegraphics[width=\columnwidth]{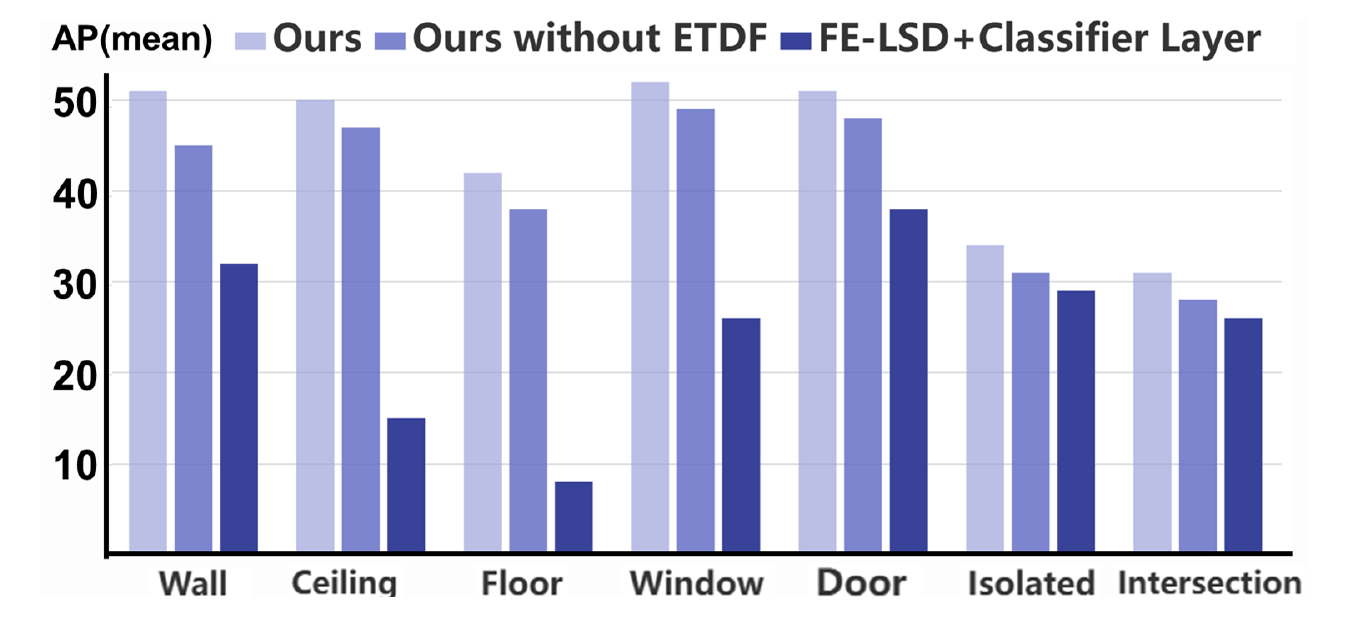}
    \end{center}
    \caption{Comparison of AP\textsuperscript{m} of event-based methods for different categories.}
    \label{fig:different_scenes}
\end{figure}

We also compared the AP$^m$ of different event-based methods under different categories, including estimating the lines of the wall, ceiling, floor, window, and door; and the estimation on isolated points and intersection points. From the results shown in Fig.~\ref{fig:different_scenes}, we can observe, that our proposed approach achieves the highest accuracy among all evaluation categories. 

To visually demonstrate the effectiveness of our proposed approach in indoor layout estimation, we show a number of estimation results of the two competing methods and annotated ground truth with different numbers of planes in an image. From Fig.~\ref{fig:image3}, it can be observed that our proposed approach can achieve stable and accurate estimations for all the scenes with different numbers of planes. The prediction accuracy of our proposed approach is significantly higher than that of FE-LSD. Additionally, even in under-exposed or over-exposed indoor scenes, our proposed approach can still achieve stable indoor layout estimation, thanks to the high dynamic range of event cameras and our proposed temporal distribution features.

\noindent \textbf{Evaluation for Different Angular Velocity.}
The maximum angular velocity of human head movement is 12.8 rad/s. As it increases, motion blur in the video becomes more pronounced. To highlight the advantages of our pipeline compared to video-based methods under high-speed motion, we classify the test dataset based on the angular velocity, from low to high, and conduct comparative tests. We implemented the frame-based indoor layout estimation method SRW-NET~\cite{srw-net}, with hyperparameters and loss functions according to its original paper. The results, shown in Fig.~\ref{fig:diff_speed}, demonstrate that as the angular velocity increases, the accuracy of our pipeline remains unaffected, while the performance of video-based layout estimation methods gradually declines due to the effects of motion blur.

Fig.~\ref{fig:diff_angular_v} illustrates the results of event-based and frame-based methods under varying angular velocities. It is evident that when the angular velocity ranges from 2 to 6 rad/s, the frame-based method, despite being affected by motion blur, is still capable of successful prediction. However, when the angular velocity exceeds 6 rad/s, the performance of the frame-based method significantly deteriorates, rendering it incapable of layout estimation. In contrast, the event-based method remains stable and effective in performing layout estimation even at angular velocities greater than 6 rad/s. This is because the event-based method can better capture scene details and dynamics under high-speed motion conditions, leading to superior performance at high angular velocities.

\subsection{Ablation Study}
\noindent \textbf{Impact of Feature Fusion Module}

As mentioned in Sec.~\ref{sec:e_fusion_module}, the Spatio-Temporal Feature Fusion Module (SFFM) allows the temporal distribution features of events to easily integrate with the backbone network of the self-attention, thus achieving the fusion of temporal and spatial features. To further validate this point, we replace the Swin-Unet part of our proposed pipeline with other self-attention-based backbone networks while keeping the rest of the pipeline unchanged. We select the most popular self-attention backbone Vision Transformer(ViT)\cite{dosovitskiy2020image} and the more complex Transformer in Transformer (TNT)\cite{han2021transformer}. 
For TNT, we integrate the event temporal distribution features with the Outer Transformer Block using SFFM. The evaluation results, shown in Table~\ref{tab:backbone}, demonstrate that, with different self-attention networks, the proposed SFFM can still successfully fuse the event temporal distribution features with the features of the backbone network itself. Meanwhile, for TNT, its $jAP^{m}$ and $sAP^{m}$ only show a slight decrease compared to Swin-Unet, maintaining a high prediction accuracy.
\begin{table}[h]
\resizebox{1.0\linewidth}{!}{
\noindent
\begin{tabular}{p{1.6cm}p{3cm}p{1.2cm}p{0.5cm}}
\toprule
\textbf{Methods} & \textbf{Backbone} & \textbf{sAP$^m$} & \textbf{jAP$^m$} \\
\midrule
\multirow{3}{*}{\textbf{Ours}}
 & Swin-Unet \cite{cao2022swin} & \textbf{48.6} & 32.3 \\
 & ViT \cite{dosovitskiy2020image} & 42.5 & 31.8 \\
 & TNT \cite{han2021transformer} & 46.3 & \textbf{35.4} \\
\bottomrule
\end{tabular}
}
\vspace{1pt}
\caption{Test results under different backbones.}
\label{tab:backbone}
\end{table}

\noindent \textbf{Impact of Event Accumulation Time Windows}
The most common approach when using event data is to set a time window, $d_t$, and accumulate the events within that fixed time window into an event frame. To evaluate whether our Ev-Layout dataset supports different sizes of time windows, we conduct tests with $d_t$ set to 1ms, 3ms, and 5ms, with the results shown in Table~\ref{tab:time_window}. As observed, our pipeline maintains high prediction accuracy across different time windows. This is primarily due to the ETDF module, which extracts features from the temporal distribution of events, allowing it to capture effective features even within very small time windows. Consequently, unlike FE-LSD, our pipeline does not exhibit a noticeable decline in accuracy as the time window decreases.

\begin{table}[h]
\centering
\resizebox{1.0\linewidth}{!}{
\setlength{\tabcolsep}{3.5pt} 
\begin{tabular}{lccc}
\toprule
\textbf{Methods} &  \textbf{Time Window} &  \textbf{sAP$^m$} &  \textbf{jAP$^m$} \\
\midrule
\multirow{3}{*}{\textbf{Ours}} 
 & 1ms &38.0&32.1\\
 & 3ms &44.5&31.5\\
 & 5ms &46.8&33.0\\
\midrule
\multirow{3}{*}{\textbf{FE-LSD+Classifier Layer}} 
 & 1ms &14.0&18.7\\
 & 3ms &20.1&25.5\\
 & 5ms &21.7&26.1\\
\midrule
\end{tabular}
} 
\caption{Test results under different time windows.}
\label{tab:time_window}
\end{table}

\section{Conclusion and Future Work}
\label{sec:conclusion}
In this paper, we introduced Ev-Layout, a comprehensive multi-modal dataset developed for indoor layout estimation and tracking on a large scale. The dataset's innovative aspects include a hybrid data collection platform that combines RGB cameras with bio-inspired event cameras, as well as the integration of time-series data from IMUs and ambient lighting conditions.

We also proposed an event-based layout estimation pipeline that includes a novel event-time distribution feature module and a spatio-temporal feature fusion module. Extensive testing shows that our method substantially enhances accuracy in dynamic indoor layout estimation compared to previous event-based approaches.

\noindent \textbf{Future Work.} Further improvements are still needed in the future. First, out of the 771,300 images in the entire dataset, only 29,000 have been annotated, thus many labeling works have to be done to improve the dataset's usability. Second, since the dataset was collected under various lighting conditions, data collected under flickering lights introduces significant noise to the event data, significantly reducing its quality. Reducing noise is key to further enhancing the dataset's usability. 

\noindent\section*{Acknowledgment}
This work is partially supported by The Natural Science Foundation of Shandong Province, Grant No. 2022HWYQ-040 and The Natural Science Foundation of Shandong Province (Major Basic Research) Grant No. ZR2024ZD12.



\begin{IEEEbiography}[{\includegraphics[width=1in,height=1.25in,clip,keepaspectratio]{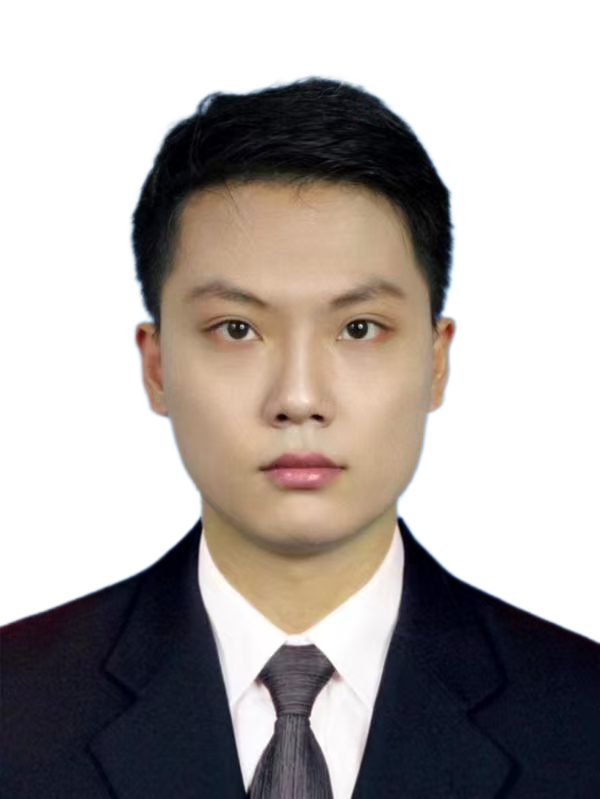}}]
{Xucheng Guo} is a PhD student at Shandong University and received his BS and MS degrees from Shandong University, China, in 2020, and Shandong University, China, in 2023, respectively. His research interests include computer vision, Virtual reality, etc.
\end{IEEEbiography}

\begin{IEEEbiography}[{\includegraphics[width=1in,height=1.25in,clip,keepaspectratio]{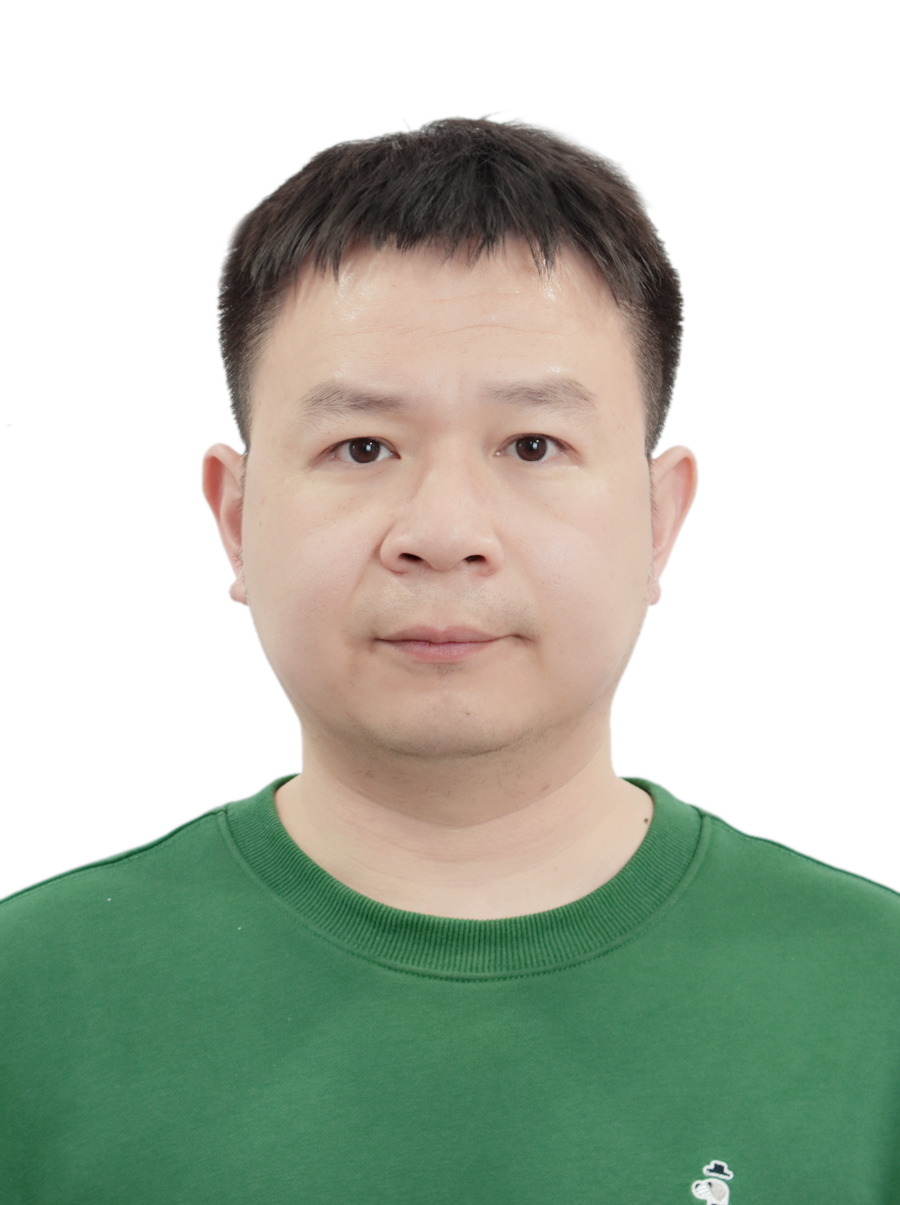}}]
{Yiran Shen} is professor in School of Software, Shandong University. He received his BE in communication engineering from Shandong University, China and his PhD degree in computer science and engineering from University of New South Wales. He published regularly at top-tier conferences and journals. Generally speaking, his research interest is sensing and computing on for immersive systems. He is a senior member of IEEE.
\end{IEEEbiography}

\begin{IEEEbiography}[{\includegraphics[width=1in,height=1.25in,clip,keepaspectratio]{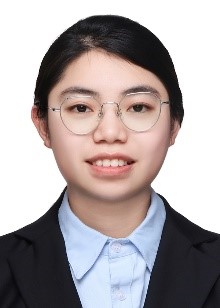}}]
{Xiaofang Xiao} is a graduate student in School of Software, Shandong University. She received her bachelor degree in software engineering from Shandong University, China in 2024. His research interests include Virtual reality, etc.
\end{IEEEbiography}

\begin{IEEEbiography}[{%
\includegraphics[width=1in,height=1.25in,clip,keepaspectratio]{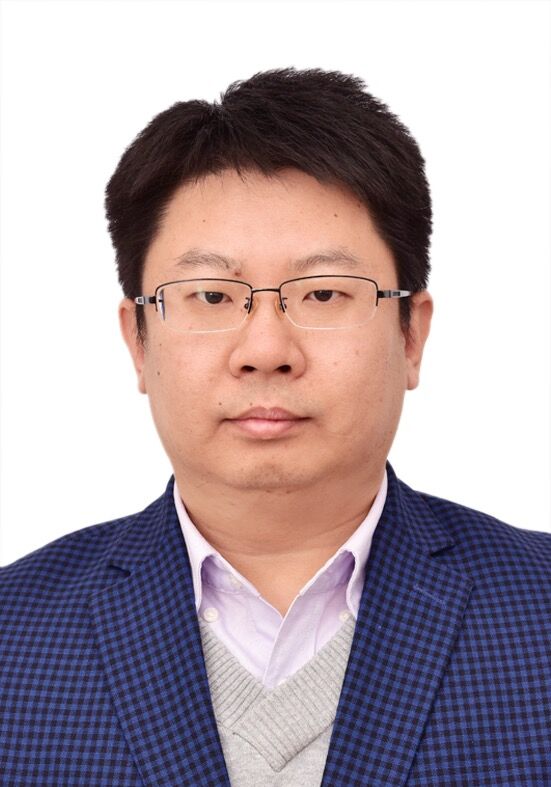}%
}]
{Yuanfeng Zhou} received his master's in 2005 and Ph.D. in 2009 from the School of Computer Science and Technology at Shandong University, Jinan, China. From 2009 to 2011, he was a postdoctoral researcher with the Graphics Group, Department of Computer Science at The University of Hong Kong. He is now a Professor at the School of Software, Shandong University, and leads the IGIP Laboratory. His research focuses on geometric modeling, information visualization, and image processing.
\end{IEEEbiography}

\begin{IEEEbiography}
[{\includegraphics[width=1in,height=1.25in,clip,keepaspectratio]{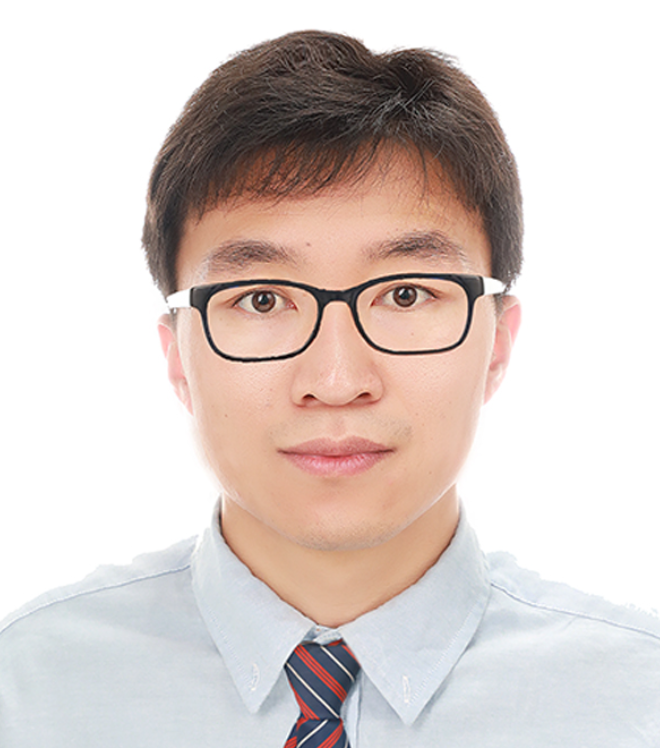}}]
{Lin Wang} (Member, IEEE) received the M.S. and
 Ph.D. (Hons.) degrees from Korea Advanced Insti
tute of Science and Technology (KAIST), Daejeon,
 South Korea, in 2017, and 2021, respectively.
 He is is currently a Tenure-track Assistant Professor at the School of Electrical and Electronic Engineering, Nanyang Technological University, Singapore. He had
 rich cross-disciplinary research experience, covering
 mechanical, industrial, and computer engineering.
 His research interests include computer and robotic vision, machine learning,
 intelligent systems (XR, vision for HCI). For more information, see https://
 vlislab22.github.io/vlislab/..
\end{IEEEbiography}

\vfill

\end{document}